\documentclass[aps,prl,reprint,superscriptaddress]{revtex4-2}

\usepackage{graphicx}
\usepackage{dcolumn}
\usepackage{bm}
\usepackage{physics}
\usepackage[english]{babel}

\begin{document}

\title{High-rate Scalable Entanglement Swapping Between Remote Entanglement Sources on Deployed New York City Fibers}

\author{Alexander N. Craddock}
\affiliation{Qunnect Inc., 141 Flushing Ave, Ste 1110, Brooklyn, NY 11205-1005}
\author{Tyler Cowan}
\affiliation{Center for Quantum Information Physics, Department of Physics, New York University, New York 10003, USA}
\author{Niccolò Bigagli}
\affiliation{Qunnect Inc., 141 Flushing Ave, Ste 1110, Brooklyn, NY 11205-1005}
\author{Suresh Yekasiri}
\affiliation{Cisco Quantum Labs, Santa Monica, CA 90404, USA}
\author{Dylan Robinson}
\affiliation{Qunnect Inc., 141 Flushing Ave, Ste 1110, Brooklyn, NY 11205-1005}
\author{Gabriel Bello Portmann}
\affiliation{Qunnect Inc., 141 Flushing Ave, Ste 1110, Brooklyn, NY 11205-1005}
\author{Aditya Verma
}
\affiliation{Qunnect Inc., 141 Flushing Ave, Ste 1110, Brooklyn, NY 11205-1005}
\author{Ziyu Guo}
\affiliation{Cisco Quantum Labs, Santa Monica, CA 90404, USA}
\author{Michael Kilzer}
\affiliation{Cisco Quantum Labs, Santa Monica, CA 90404, USA}
\author{Jiapeng Zhao}
\affiliation{Cisco Quantum Labs, Santa Monica, CA 90404, USA}
\author{Mael Flament}
\affiliation{Qunnect Inc., 141 Flushing Ave, Ste 1110, Brooklyn, NY 11205-1005}
\author{Javad Shabani}
\affiliation{Center for Quantum Information Physics, Department of Physics, New York University, New York 10003, USA}
\author{Reza Nejabati}
\affiliation{Cisco Quantum Labs, Santa Monica, CA 90404, USA}
\author{Mehdi Namazi}
\email{mehdi@quconn.com}
\affiliation{Qunnect Inc., 141 Flushing Ave, Ste 1110, Brooklyn, NY 11205-1005}

\date{\today}

\begin{abstract}
Entanglement swapping between photon pairs generated at physically separated nodes over telecommunication fiber infrastructure is an essential step towards the quantum internet, enabling applications such as quantum repeaters, blind quantum computing, distributed quantum computing, and distributed quantum sensing. 
However, successful networked entanglement swapping relies on generating indistinguishable pairs of photons and preserving them over deployed fibers.
This has limited most previous demonstrations to laboratory settings or relied on sophisticated methods to maintain the necessary indistinguishability. 
Here, we demonstrate a scalable entanglement swapping experiment using naturally indistinguishable entanglement sources based on warm atomic vapor cells. 
Without sharing lasers or optical frequency references between nodes, nor the need for pulsing the sources, we achieve a swapping rate of nearly 500 pairs/s while maintaining the CHSH parameter above 2. Additionally, we demonstrate the scalability of our method by maintaining the quality of the entanglement swapping on $17.6$ km of deployed fibers in NYC, relying on commercially available SPADs at the spoke nodes, SNSPDs at the hub and standard time‑synchronization techniques. 
Our work paves the way for the practical deployment of large-scale hub-and-spoke quantum networks within cities and data centers.

\end{abstract}

\maketitle


\section{\label{sec:level1}Introduction }

Entanglement swapping is an important operation in the quantum internet \cite{Kimble2008-fo}, critical in use cases such as quantum repeaters \cite{Azuma2023-dl}, distributed quantum computing \cite{Monroe2014-ja,Pirandola2016-id,Wehner2018-rz}, and quantum sensing \cite{Khabiboulline2019-jz}. All such use cases require the entanglement swapping process to be high rate, stable, scalable, and deployable across telecommunication networks or within buildings, such as quantum data centers (QDCs) \cite{Shchukin2022-bl}.

More specifically, entanglement swapping between remote photonic entanglement sources is not only crucial for creating long distance quantum networks \cite{Kaltenbaek2009-ax,Sangouard2008-qo} but is also required for blind quantum computing \cite{Broadbent2009-je,Li2014-gt} and enhancing the entanglement rate between quantum processors over a distributed quantum computing network \cite{shapourian2025quantum}. 
However, while many groups have reported progress towards this goal in recent years, previous demonstrations have proven that efficient entanglement swapping between independent bi-photon sources is a very hard task to achieve \cite{Davis2025-et,Sun2017-vj}. In addition, the deployment of such networks on the telecom infrastructure adds significant practical challenges.

\begin{figure*}[t]
    \centering
    \includegraphics[width=\linewidth]{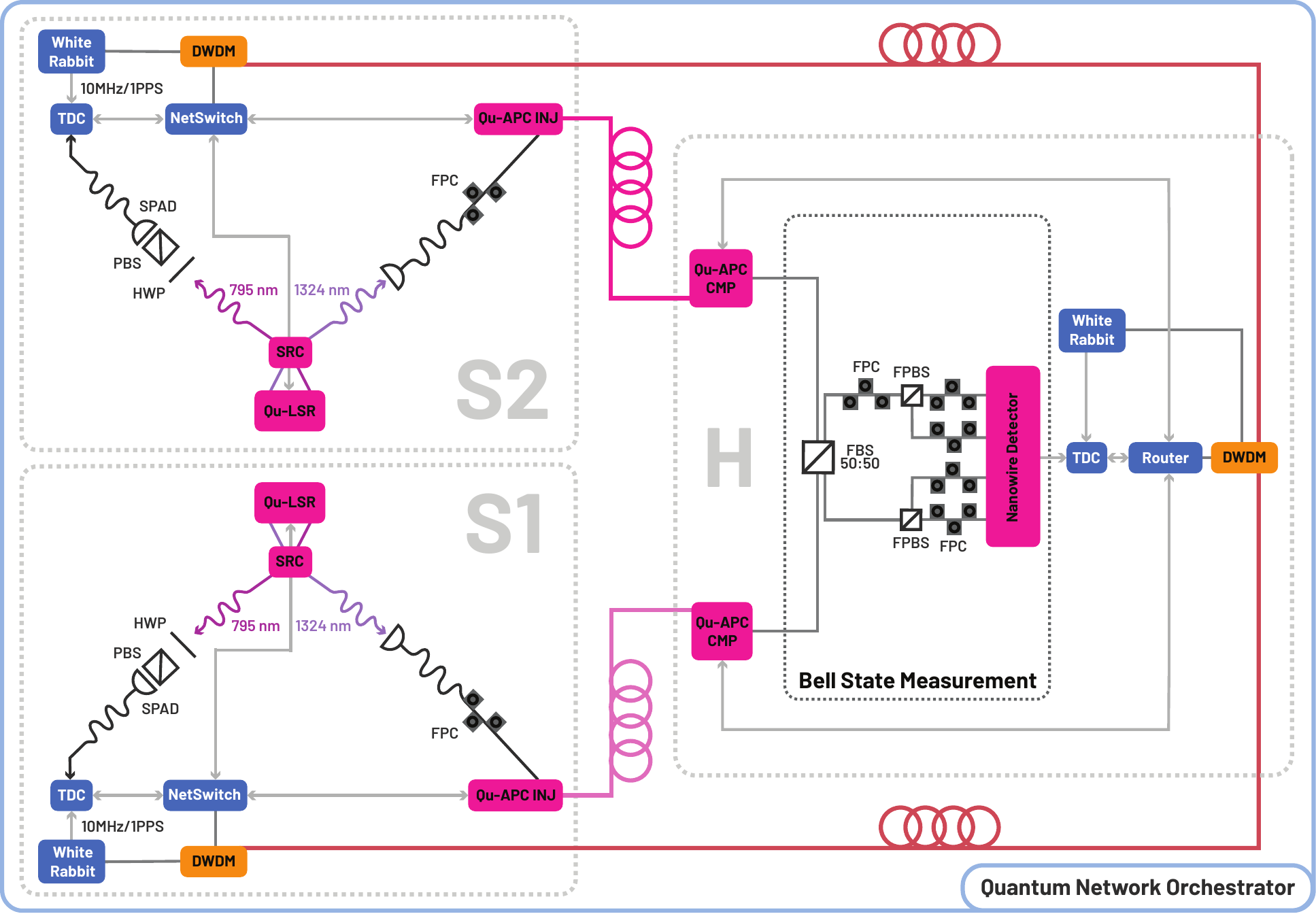}
        \caption{Experimental layout of a three-node entanglement swapping experiment. Nodes S1 and S2 host the independent entanglement sources while H is where the entanglement swapping happens. Qu-LSR: Qunnect's Dual laser systems for the bi-chromatic sources. SRC: Atomic vapor cell sources. HWP: Half waveplate. SPAD: Single photon avalanche detector. TDC: Time-to-digital converter (time tagging device). Qu-APC: Qunnect's automated polarization compensating devices (INJ: Injector. CMP: Compensator). FBS: Fiber beamsplitter. FPBS: Fiber polarization beamsplitter . All three nodes are controlled by Cisco’s centralized Quantum Network Orchestrator.  }
    \label{fig:experimental_diagram}
\end{figure*}

One major challenge with swapping is the requirement of indistinguishability \cite{Basso_Basset2019-qh}. 
For this process to be successful, not only must the photons from different sources be identical to one another, but fiber effects, such as polarization drifts, must be mitigated to ensure high fidelity entanglement swapping.
With sources such as Spontaneous Parametric Down-Conversion (SPDC) \cite{Pan1998-xy}, color centers in diamonds \cite{Stolk2024-hh}, and quantum dots \cite{Basso_Basset2019-qh}, producing identical sources is a challenge.
For sources that produce naturally indistinguishable photons, such as atomic and ionic systems, preserving the degrees of freedom of the photons across the network can be non-trivial \cite{Stolk2024-hh,Liu2026-kw}. 
For instance, while polarization qubits are widely used in the world of quantum information, to date there has never been a demonstration to show polarization entanglement swapping between fiber-gapped sources.

Another challenge with entanglement swapping is achieving high rates.
Since the operation relies on the simultaneous arrival of photons from two different sources, probabilistic entanglement sources need to have high spectral brightness. 
To date, most experiments using such sources have resulted in rates below 1 event/s \cite{Davis2025-et,Park2020-kx,Sun2017-vj}.
There are some exceptions, with the highest rate reported at 108 pairs/s \cite{Jin2015-lg}.
Although impressive, this setup relied on two SPDC sources being pumped using the same laser, making it impractical to implement over a network.
Deterministic sources, such as trapped ions or neutral atoms, do not necessarily have this timing issue because the photons can be made to arrive simultaneously.
However, there are often issues with the collection efficiency of the photons and a lack of native telecom compatibility, which limits the distance over which swapping can be performed.
Here, the highest rates achieved using such platforms reach a rate of 182 events/s at 422-nm, separated by roughly 2 meters \cite{Stephenson2020-al}.

Here, we demonstrate entanglement swapping with polarization qubits that not only enables high‑rate swapping but is also highly scalable for larger spoke and hub models, such as in QDCs and metropolitan networks. 
Using two fully independent, highly bi-chromatic atomic vapor based entanglement sources, we achieve a swapping rate exceeding 470 pairs/s when the sources are ``local" to the hub (all within the same building).
The swapping is performed at telecom wavelengths and occurs at a superconducting nanowire equipped hub, whereas the entangled photons are in the near infrared and detected with lower cost single photon avalanche detectors (SPADs).
This architecture allows for the easy addition of more spokes, given the low size, weight, power, and cost requirements of the spoke equipment.
Additionally, we demonstrate the scalability of our platform by performing the swapping experiment on 17.6-km of deployed fibers in New York City, placing the hub at a commercial data center, QTD Systems. 
Using this layout, we achieve the first Hz-level polarization entanglement swapping over a commercially deployed network.

\section{Experimental Details}

Fig.~\ref{fig:experimental_diagram} shows a hardware schematic for the two spokes (S1 and S2) and the hub (H) for the following experiments.
The two spokes have identical layouts and are designed to be fully operational at room-temperature.
In contrast, the hub is equipped with superconducting nanowire detectors (SNSPDs) and therefore requires cryogenic cooling.
This layout allows for a low-cost, low-maintenance expansion of the number of spoke nodes, all connected to the same central hub, without the need for any additional cryogenic SNSPDs.

\begin{figure*}[t]
    \centering
    \includegraphics[width=\linewidth]{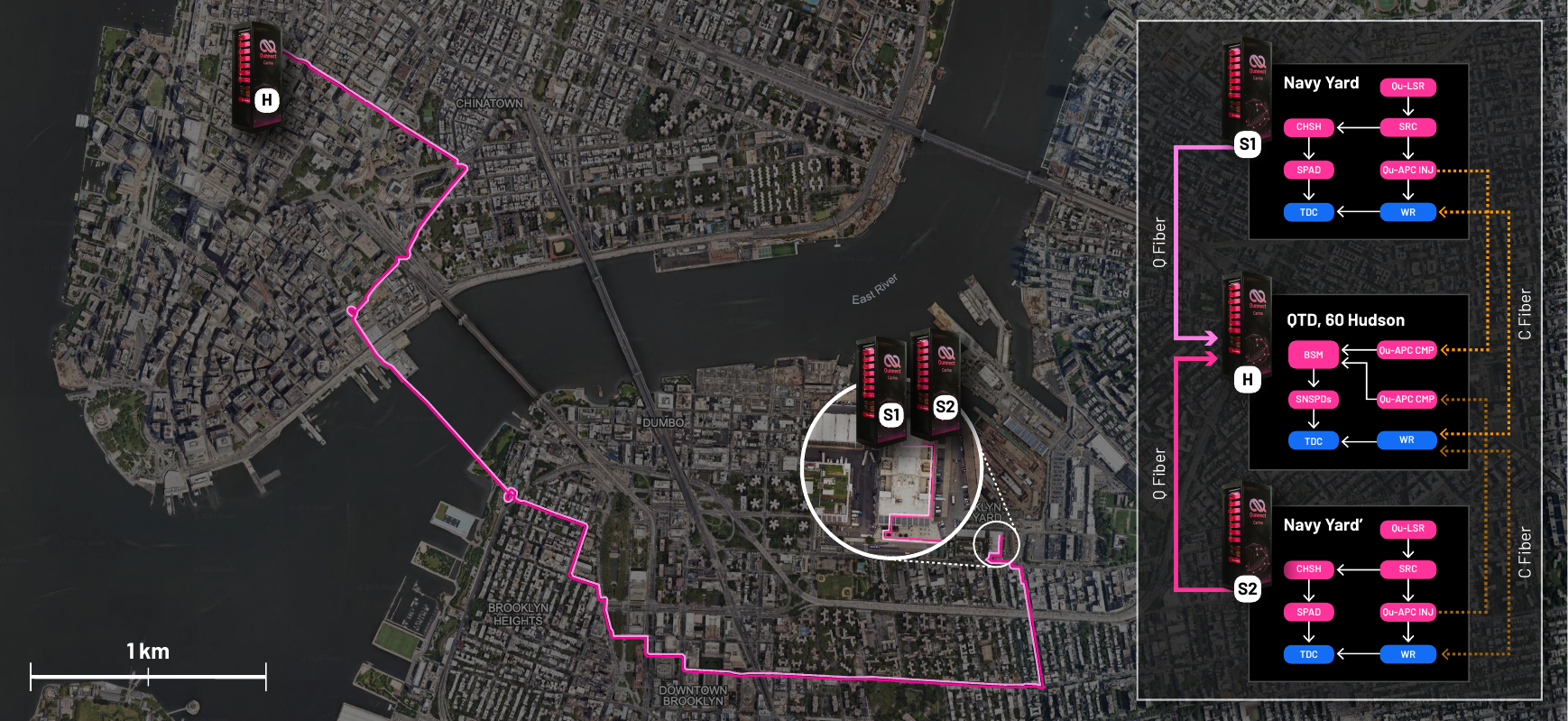}
    \caption{Path of fibers for network experiments.
    Inset shows hardware used at the three locations.
    WR: White Rabbit, TDC: time tagging device, SPAD: single photon avalanche detector, SNSPD: superconducting nanowire single photon detector, Qu-LSR: Qunnect dual‑laser pump system, SRC: entanglement source, CHSH: apparatus for performing CHSH type measurements, SWAP: Bell state measurement optics for swapping, Qu-APC: Qunnect's automated polarization compensating device.}
    \label{fig:GQmap}
\end{figure*}

At the spokes, there are fully independent entanglement sources based on spontaneous  four-wave mixing in a warm rubidium vapor.
Using a dual‑laser pump system (Qu‑LSR, Qunnect) to drive the vapor with $780$-nm and $1367$-nm pumps, the source generates bi-chromatic pairs of $795$-nm and $1324$-nm photons that are entangled in a $\ket{\Phi_+}=\frac{1}{\sqrt{2}}\left(\ket{HH}+\ket{VV}\right)$ Bell state \cite{craddock2023highrate}.
In contrast to previous work, here the entanglement source utilizes an enriched rubidium 85 vapor as opposed to rubidium 87.
This was done to provide biphotons that are more temporally compatible with off-the-shelf SPADs.

From the entanglement sources, the $795$-nm photons pass through a free-space half-waveplate (HWP) followed by a polarizer, allowing us to perform CHSH-like measurements between the two spokes. 
Post-polarizer, $795$-nm photons are coupled into fiber and directed towards a SPAD (detection efficiency $\approx 65$\%).

\begin{figure*}[t]
    \centering
    \includegraphics[width=\linewidth]{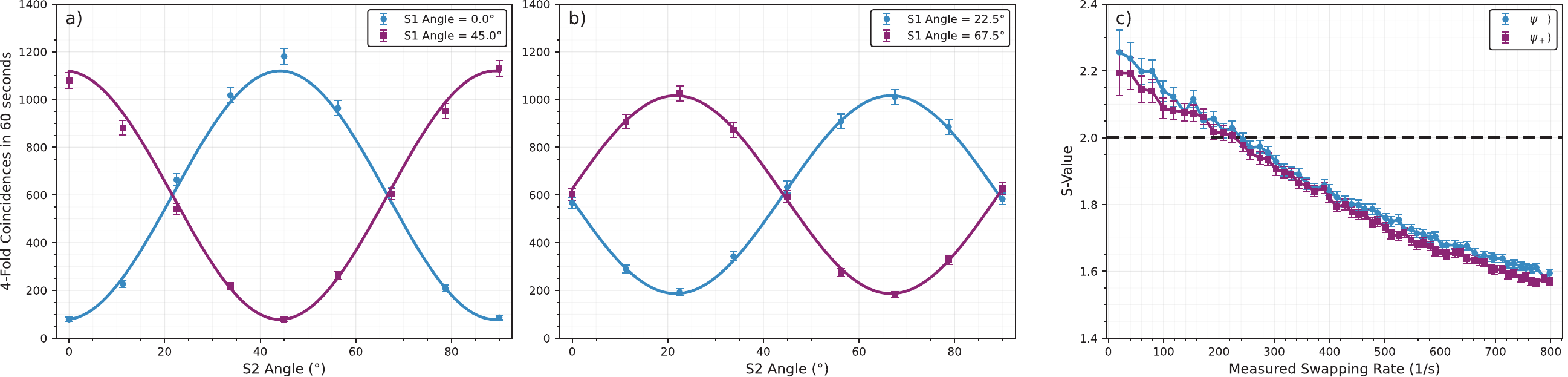}
    \caption{Experimental results for the ``local" swapping experiment.
    a) and b) show interference fringes as the HWP at S1 and S2 are rotated.
    From the data we extract Bell's S parameter and the swapping rate.
    We do this many times, changing the region of interest used for analysis to obtain the plot in c) of the S parameter as a function of the swapping rate.}
    \label{fig:swaplocal}
\end{figure*}

The $1324$-nm photons are coupled into the fiber and directed towards the hub, where swapping occurs.
At the hub, the $1324$-nm photons from the two spokes are passed through a fiber-based Bell state measurement (BSM).
This consists of a fiber-based 50:50 splitter followed by a pair of fiber-based polarizing beamsplitter cubes (PBS).
We use a set of paddles between the 50:50 splitter and one of the PBS to allow us to orient the axes of the two PBS in the BSM to be the same.
The four output fibers of the BSM are coupled to the hub's SNSPDs ($>80$\% and $>93$\% detection efficiency for detectors when the hub is local and remote, respectively).
The excess loss through the BSM setup is $\approx 1.3$ dB and $\approx 1.9$ for the ``local" and networked experiments, respectively. 
The difference is related to some additional fiber optics that were required for remote control in the remote case and could be neglected in the future.

For the ``local" experiments, the fibers used are relatively short and experience little polarization drift.
As a result, active polarization compensation is unnecessary, so the automated polarization compensator (Qu-APC) is bypassed.
For the city-wide experiment, we see fiber drift similar to previous experiments \cite{Craddock2024-ci,Sena2025-im,Shi2026-eu}, and we similarly use the Qu-APC injector-compensator pair to stabilize the fibers to the hub.
These devices operate as a pair and use time-division multiplexing, switching between the $1324$-nm photons from the source and a classical laser at a similar wavelength to monitor the fiber and compensate for any polarization drifts that may have occurred.
For the experiments shown, the Qu-APC was operated to check for polarization drifts every thirty seconds.
When used, the uptime of the Qu-APC was in excess of $99$\% and added an insertion loss of $\approx2$ dB per link.
In both experiments, a set of paddles is used on the fibers that go to the hub.
These are used to correct for the polarization rotation experienced by the photons traveling down the fiber but are not adjusted during the experiments.

For both sets of experiments, data is collected from the various detectors using time tagging devices local to those detectors.
A classical network, utilizing a separate set of fibers from the ones that transmit the $1324$-nm photons, allows data to be collated between the three locations using TCP/IP.
We use the White Rabbit protocol to synchronize the time taggers at the three nodes, with the White Rabbit devices operating over the same fibers as the TCP/IP network.
Cisco's Quantum Network Orchestrator facilitates network operation by centrally controlling three experimental nodes. 
The integration of such an orchestrating software suite with the quantum layer is an essential step towards the expansion of such demonstrations to many-node configurations.

For both demonstrations, S1 and S2 are in two separate rooms (separated by $\approx30$ meters) of the same building at the Brooklyn Navy Yard in Brooklyn, New York.
As mentioned above, the two rooms are fully fiber-gapped from each other.
For the ``local" experiments, the hub is co-located with S1.
As such, the fiber loss from S1 to the hub is minimal, while the fiber loss from S2 to the hub is $\approx 1$ dB.

For city-wide experiments, the hub is located in the historic 60 Hudson building in the QTD systems datacenter.
The spokes are connected to the hub by fibers, shown in Fig.\ref{fig:GQmap}, of length $\approx 8.8$ km, with a loss of $\approx 5$ dB (measured by optical time‑domain reflectometry at 1310 nm; the loss at 1324 nm is similar), bringing the combined fiber distance between S1 and S2 to $\approx 17.6$ km.

\section{Local Entanglement Swapping}

As a benchmark of the swapping protocol between the independent sources, we first run the experiment with all the nodes located within the same building.
Both entanglement sources are operated at a point where the peak cross-correlation function between the $795$ and $1324$-nm photons is $\approx 80$.
Here, the measured pair rate of the two sources at the detectors is $\approx 1.7$ M/s and $\approx 1.3$ M/s for S1 and S2 respectively, corresponding to a pair rate from the source of $\approx 4.1$ and $4.0$ M/s (the difference is attributed to slight fiber coupling differences between the sources).
We optimize the paddles going from both S1 and S2 to the hub to ensure that the $1324$-nm photons arriving at the hub's BSM are still in the state $\ket{\Phi_+}=\frac{1}{\sqrt{2}}\left(\ket{HH}+\ket{VV}\right)$ with the $795$-nm photons at the spokes.
For each set of angles for the spoke HWP's we record the relevant four-fold coincidences for sixty seconds.

Fig.~\ref{fig:swaplocal} a) and b) show the four-fold coincidences vary as a function of the waveplate angles at S1 and S2 for one of the heralded Bell states.
We observe the interference fringes typical of a polarization entangled pair of photons.
We note here that the differing visibility for the curves in a) and b) is expected for our system.
For a), the visibility is limited only by the effective cross-correlation (the cross-correlation averaged over the region of interest used for coincidence measurements) of the entanglement sources.
For b), the visibility is limited by the cross-correlation of the sources and the Hong-Ou-Mandel (HOM) visibility of the sources ($\approx 80$\% for the data shown in Fig.~\ref{fig:swaplocal}).

In Fig.~\ref{fig:swaplocal} c) we re-analyze the collected data using different regions of interest for the four-fold coincidence.
We show the relationship between the swapping fidelity (using the Bell S-parameter as a proxy) and the measured swapping rate.
As can be seen $S>2$ for the measured swapping rates exceeding $200$ /s. 
Accounting for the SPADs detection efficiency of 65\%, our measurement corresponds to an entanglement swapping rate in excess of $470$ /s.
This is to the best of our knowledge the highest rate of swapping achieved between thermal entanglement sources, regardless of the qubit type. 
These results are also nearly four orders of magnitude better than the previous state of the art between two independent entanglement sources \cite{Park2020-kx}. 
Lastly, to the best of our knowledge, these results mark a few fold improvement in rate compared to the state of telecom swapping between deterministic sources such as ion-traps or SiV defects, paving the path towards hybrid distributed quantum computing systems with high rate entanglement generation. 

\section{Entanglement Swapping Over Deployed NYC Fibers}

\begin{figure}[t]
    \centering
    \includegraphics[width=\linewidth]{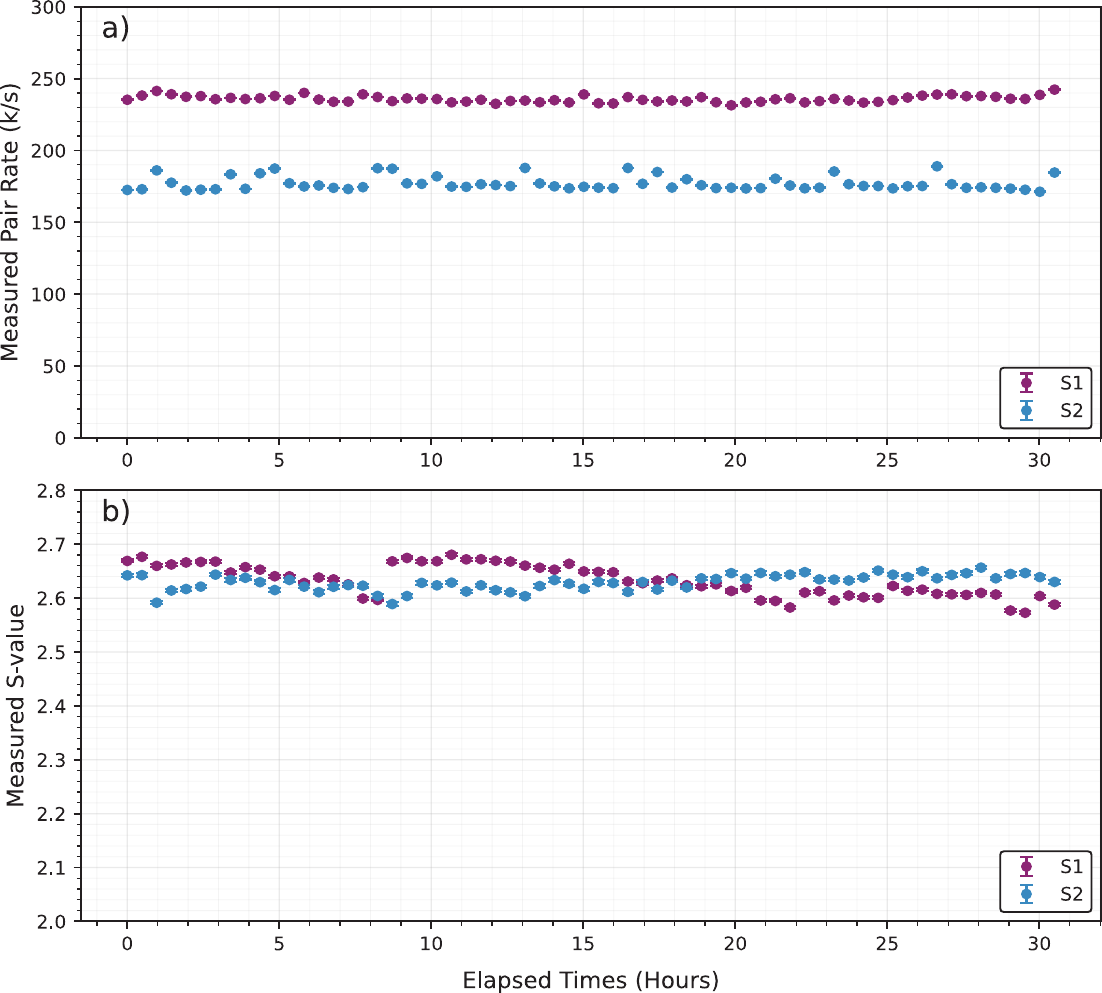}
    \caption{Single spoke entanglement distribution stability results.
    a) shows the measured pair rate and b) shows the entanglment fidelity, parametrized by the Bell's S parameter, as a function of time.}
    \label{fig:remote_stability}
\end{figure}

Next, we demonstrate entanglement swapping over deployed fiber in New York City.
As discussed, here the hub is connected to the spokes by $\approx 8.8$ km of fiber, with the fiber route shown in Fig.\ref{fig:GQmap}.
We operate the sources in the same configuration used for the local experiments, where the peak measured cross-correlation is $\approx 80$.
Due to the increased losses from the spokes to the hub, we see lower pair rates of $\approx 240$ k/s and $\approx 180$ k/s for S1 and S2, respectively.
Given that the sources operate nearly identically to the ``local" experiment, we estimate an additional loss of $8.3$ and $8.6$ dB for S1 and S2.
Taking into account the difference in detection efficiency between the experiments and the differing loss in the BSM (used in these measurements), this corresponds to a link loss of $\approx 8.2$ and $\approx 9.5$ dB for the S1-hub and S2-hub links, respectively.
We independently measure losses of $\approx 7$ dB for the deployed fiber ($\approx 5$ dB for the fiber itself and $\approx 2$ dB for the Qu-APC), and we attribute the additional loss to those incurred in connecting the equipment at the spokes and hub to the deployed fiber due to patch panels and connectors.

For this experiment, we again optimize the paddles so that the $1324$ and $795$-nm photons of both spokes are in the $\ket{\Phi_+}=\frac{1}{\sqrt{2}}\left(\ket{HH}+\ket{VV}\right)$ state after fiber propagation.
We operate the Qu-APC to check the fiber drift every thirty seconds and perform corrections if necessary.

Before performing the swapping experiment, we first check the stability of the single fiber entanglement distribution.
As shown in Fig.\ref{fig:remote_stability}a) the measured pair rate through the network is stable to $\pm 5$\% over 30 hours.
More importantly than the pair rate, for entanglement swapping over the deployed fiber to be successful, we require that the fidelity of the entanglement generated at the two spokes is preserved upon arriving at the hub.
To this end, we also monitor the Bell S-parameter over the same 30-hour period for the two individual sources and observe relatively stable behavior, as demonstrated in Fig.\ref{fig:remote_stability}b). 
We note that the Qu-APC has some programmed tolerance threshold for allowable fiber drift before it compensates. Therefore, some small drift, such as that shown in the figure, is expected.

\begin{figure*}[t]
    \centering
    \includegraphics[width=\linewidth]{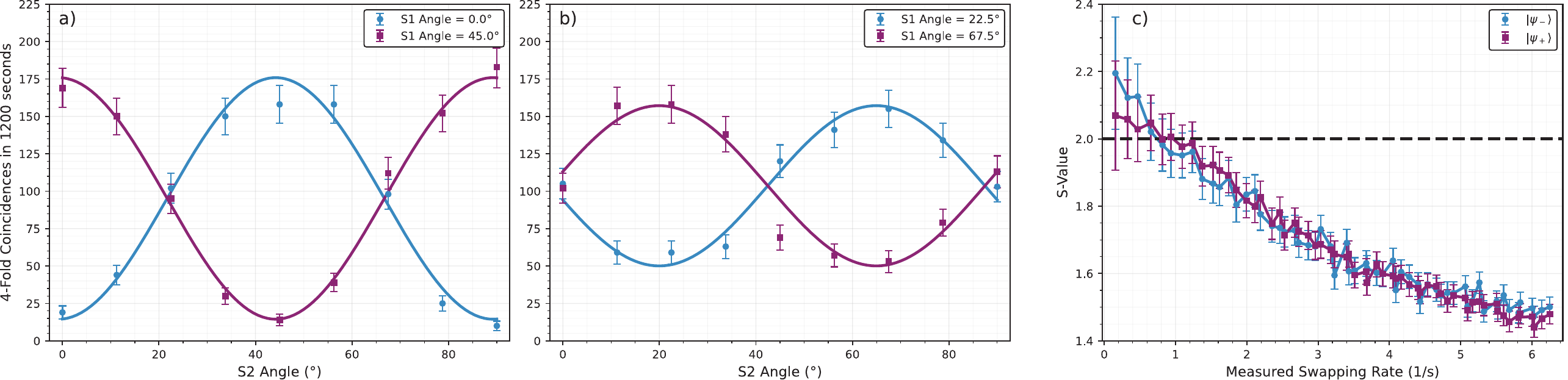}
    \caption{Experimental results for the network swapping experiment.
    a) and b) show interference fringes as the HWP at S1 and S2 are rotated.
    From the data we extract Bell's S parameter and the swapping rate.
    We do this many times, changing the region of interest used for analysis to obtain the plot in c) of the S parameter as a function of the swapping rate.}
    \label{fig:swapRemote}
\end{figure*}

Having established the stability of the network, we move on to demonstrate entanglement swapping using the deployed fibers.
As in the local experiment, we again set the HWPs at the two spokes to defined angles before accumulating data for sixty seconds.
We cycle through each set of HWP angles twenty times so that each angle pair has data recorded for $1200$ seconds.
Although we have seen that the entanglement distribution through the network is relatively stable, we perform this method of data accumulation to spread the effect of any small fluctuations in either source rate or fiber polarization drift over all the angle pairs, so as not to bias the results.
As we rotate the HWPs we again observe characteristic interference fringes consistent with polarization entangled photon pairs, as shown in Fig.\ref{fig:swapRemote} a) and b).

Similarly to the local data, we re-analyze the swapping data using different regions of interest to examine the relationship between the swapping rate and fidelity, the results of which are shown in Fig.\ref{fig:swapRemote}c).
We observe values of $S>2$ for measured swapping rates over $0.65$ /s.
Accounting for the SPADs detection efficiency of 65\%, our measurement corresponds to an entanglement swapping rate in excess of $1.5$ /s.
To the best of our knowledge, this is the first time polarization entanglement swapping has been demonstrated over deployed fiber. 
Additionally, this experiment marks the highest swapping rate achieved between any two quantum platforms across a deployed network. 

\section{Discussion and Conclusion}

High-rate, field-deployable entanglement swapping is an important step in the development of the quantum internet. 
Here, we present experimental data from our testbed, GothamQ, in New York City, demonstrating the first city-wide polarization entanglement swapping experiment while achieving the highest swapping rates to date between any two quantum systems.
More importantly, we deploy a scalable ``Spoke-and-Hub" network in which the spokes do not require any sophisticated or cryogenically cooled technology, including the single photon detectors.
Such an architecture is essential for scaling networks within cities or QDCs.

In most swapping experiments with paired sources, the photons used are near degenerate.
Therefore, we emphasize the use of atomic-based bi-chromatic entanglement sources in this work.  
Not only do these sources eliminate the need for using cryogenic detectors or input laser sharing, they also remove the need for frequency conversion in many quantum repeating and distributed protocols. 
The newly entangled 795-nm (or 780-nm) photons over the telecommunication network are natively compatible with a large variety of quantum memories, neutral atom quantum computers, atomic clocks, and quantum sensors.

Lastly, large-scale quantum networks require both scalable hardware and software to work seamlessly together.  
Bringing together Qunnect's Carina system with Cisco's orchestrator software for a quantum networking operation at one of Manhattan's busiest data centers aims to demonstrate the readiness of quantum entanglement networks for industrial deployments beyond point-to-point use-cases.

\begin{acknowledgments}
\section{acknowledgments}
We would like to sincerely thank QTD Systems LLC, especially Peter Feldman, for not only accommodating our hardware at their datacenter but also for providing seamless access and the partnership necessary for a first of its kind quantum experiment. Additionally, a major thank you to Vijoy Pandey and Ramana Kompella of Cisco and Noel Goddard of Qunnect, without whom this collaboration would not have been possible.  
\end{acknowledgments}

\nocite{*}

\bibliography{main}

\end{document}